# Tuning of the ultrafast demagnetization by ultrashort spin polarized currents in multi-sublattice ferrimagnets.

**Deeksha Gupta**[1], **Maryna Pankratova**[2,6], **Matthias Riepp**[1], **Manuel Pereiro**[2], **Biplab Sanyal**[2], **Soheil Ershadrad**[2], **Michel Hehn**[3], **Niko Pontius**[4], **Christian Schüßler-Langeheine**[4], **Radu Abrudan**[4], **Nicolas Bergeard**[1], **Anders Bergman**[2], **Olle Eriksson**[2,5], **and Christine Boeglin**[1*]

[1]Institut de Physique et de Chimie des Matériaux de Strasbourg, UMR7504, CNRS et Université de Strasbourg, 67034 Strasbourg, France

[2]Department of Physics and Astronomy, Uppsala University, Box 516, SE-75120 Uppsala, Sweden

[3]Institut Jean Lamour, Université Henri Poincaré, Nancy, France

[4]Helmholtz-Zentrum Berlin für Materialien und Energie GmbH, Albert-Einstein Str. 15, 12489 Berlin, Germany

[5]Wallenberg Initiative Materials Science for Sustainability (WISE), Uppsala University, Box 516, SE-75120 Uppsala, Sweden

[6]Department of Engineering Sciences, University of Skövde, SE-541 28 Skövde, Sweden

[*]christine.boeglin@ipcms.unistra.fr

# ABSTRACT

Femtosecond laser pulses can be used to induce ultrafast changes of the magnetization in magnetic materials. Several microscopic mechanisms have been proposed to explain the observations, including the transport of ultrashort spin-polarized hot-electrons (SPHE). Such ultrafast spin currents find growing interest because of the recent challenges in ultrafast spintronics, however they are only poorly characterized. One of the key challenges is to characterize the spin-polarized ultrafast currents and the microscopic mechanisms behind SPHE induced manipulation of the magnetization, especially in the case of technologically relevant ferrimagnetic alloys. Here, we have used a combined approach using time- and element-resolved X-ray magnetic circular dichroism and theoretical calculations based on atomistic spin-dynamics simulations to address the ultrafast transfer of the angular momentum from spin-polarized currents into ferrimagnetic $Fe_{74}Gd_{26}$ films and the concomitant reduction of sub-lattice magnetization. Our study shows that using a Co/Pt multilayer as a polarizer in a spin-valve structure, the SPHE drives the demagnetization of the two sub-lattices of the $Fe_{74}Gd_{26}$ film. This behaviour is explained based on two physical mechanisms, i.e., spin transfer torque and thermal fluctuations induced by the SPHE. We provide a quantitative description of the heat transfer of the ultrashort SPHE pulse to the $Fe_{74}Gd_{26}$ films, as well as the degree of spin-polarization of the SPHE current density responsible for the observed magnetization dynamics. Our work finally characterizes the spin-polarization of the SPHEs revealing unexpected opposite spin polarization to the Co magnetization, explaining our experimental results.

## Introduction

Ferrimagnets are important materials in order to push spintronics and magnetic data storages towards the subpicosecond regime while ensuring low energy consumption. These materials are also among systems that show single pulse all optical switching, an important property for applications using ultrafast spintronics[1–5]. Since more than 25 years, femtosecond laser pulses have been used as an ultrafast source of excitation to induce ultrafast changes in the magnetization, motivating many experimental and theoretical descriptions[4,6–14]. In order to manipulate the magnetization in ferro- and ferrimagnetic films, ultrashort spin current pulses have recently been revealed to be an extremely promising way for applications in ultrafast spintronics since they are known to launch ultrafast spin dynamics in different magnetic films with low heat dissipation[15–24]. Such ultrafast spin currents can be launched using femtosecond laser pulses exciting ultrathin metallic films. This method of generating ultrashort currents has attracted increasing technological interest related to their ultrashort duration of a few hundred femtoseconds, which is compatible with subpicosecond manipulations of magnetization.

In previous works, ultrashort current pulses have been shown to flow between two ultrathin ferro- or ferrimagnetic layers[19,25,26], and they are able to produce subpicosecond demagnetization or picosecond switching without any external magnetic fields. Recently, specifically grown spin-valve structures with two separated magnetic layers have been used to produce spin-polarized hot-electron (SPHE) currents in a hard-magnetic layer. A second free (soft) magnetic layer is used to observe the induced dynamics. Several characteristic features can be extracted from the literature; for instance, the antiparallel orientation of both magnetic layers in the spin valve favors the hot-electron (HE) induced switching of the soft layer[19,24]. Surprisingly, recent counterintuitive results revealed that other spin valve systems show switching in the soft layer only for a parallel orientation of the hard and soft layer[19]. Those results showed that the orientation for successful switching depends on fluence, which raises questions about the multiple mechanisms behind spin switching.

These different results from the literature ask for a theoretical microscopic description of the mechanisms involved in the transfer of spin angular momentum between the SPHE current and the magnetic moments in ferro- and ferrimagnets. One of the first combined models one can think of to describe ultrafast demagnetization associates thermal induced reduction of the magnetic order with non-thermal angular momentum transfer. In multi-sublattice ferrimagnets, the role of spin current transfer is even more complex[27] since additional local mechanisms may emerge. For instance, local ultrafast transfer of angular momentum between two exchange coupled sublattices, suggested by Mentink et al.[28], has been confirmed by experimental observations[29]. It evidences that during the ultrafast loss of magnetization in each of the sub-systems, the total angular momentum is conserved over a few hundred femtoseconds, involving two compensating angular momenta, which flow in opposite directions. The situation is drastically different and is expected to be more complex by using an excitation source of spin polarized currents with a pulse duration of a few hundred of femtoseconds, potentially transferring angular moment to the ferrimagnet during the first hundreds of femtoseconds.

The microscopic processes defining the ultrafast excitation by ultrashort spin currents of such complex multi-sublattice ferrimagnets thus need a detailed and microscopic understanding in the subpicosecond time scale. Among others, the information about the pulse energy density (or fluence), spin polarization, and pulse duration of the SPHE pulses is essential to describe the ultrafast excitations. It is indeed well documented that the pulse duration together with the pulse fluence are important parameters driving the ultrafast demagnetization and switching when excited by infrared (IR) or hot electrons (HE)[30,31]. When using IR pump pulses, the duration of the pulse is well controlled and can be tuned to reach the desired thresholds (100 fs - 2 ps). The situation is more complex for ultrashort spin current pulses where the pulse durations are produced via diffusion and propagation through the films, leading to only indirect characterization of the pump duration and spin polarization. We estimate that the slightly longer pulses of HE of a few hundred of femtoseconds will not be detrimental to the angular moment transfer in ferrimagnets because the demagnetization dynamics measured in those materials are much longer and last for almost 1 ps.

We report here on combined experimental and theoretical results evidencing the spin dependent hot electron (SPHE) induced demagnetization on the ultrafast time scale on $Fe_{74}Gd_{26}$ alloy in a specifically designed spin valve structure. We provide different quantitative numbers for pulse duration and spin polarization that characterize the hot-electron pulses. The experimental results were obtained by time-resolved X-ray magnetic circular dichroism (TR-XMCD) at the transition-metal (TM) $L_3$ and rare-earth (RE) $M_5$ edges at the BESSY II Femtoslicing source of the Helmholtz-Zentrum Berlin[7,32,33]. This experimental method combines element and magnetic sensitivity[34–36] with femtosecond time resolution, resolving the ultrafast magnetization dynamics in ferrimagnetic $Fe_{74}Gd_{26}$ alloys. Here, we give the first experimental results evidencing the timescales of the ultrafast quenching of the magnetization in ferrimagnetic $Fe_{74}Gd_{26}$, induced by ultrashort pulses of spin currents as produced in a collinear magnetic spin valve structure (CoPt / Cu / $Fe_{74}Gd_{26}$). In addition, it is reported here that, relying on experimentally defined geometry and composition as well as interatomic exchange, theoretical modeling based on atomistic spin-dynamics simulations reproduces the experimental ultrafast dynamics of this system. This fact allows to identify the microscopic process of spin angular momentum transfer at the shortest time scale[14,37]. The proposed model is a combination of two processes: one is thermal in origin due to HE induced heating of the spin system, and the second is a non-thermal spin transfer torque (STT) resulting from spin angular momentum transfer. Note that it is out of the scope of this paper to explicitly model the SPHE current generated from the Co/Pt multilayer. Currently, the physics governing the generation and transport is still under debate, and it is still speculated that multiple mechanisms are possible to generate such currents: hot electron spin filter effect through electron scattering, superdiffusive spin transport and magnon excitations via the ultrafast loss of angular momentum in the ultrathin Co/Pt multilayer[12,13,15,17,38,39].

As detailed below, the theoretical model employed here reproduces the experimental data and reveals how SPHE excitations drive the demagnetization in both sub-lattices of $Fe_{74}Gd_{26}$. Most noteworthy, the theoretical calculations reproduce the time scales and amplitudes of the experimental results recorded at Fe $L_3$ and Gd $M_5$ edges, leading to an indirect determination of the sign of spin polarization in the SPHE current. This model can further be used to predict the impact of fluence dependent variations of the HE induced heating and of STT in both of the $Fe_{74}Gd_{26}$ sub-lattices.

## Results

**Experimental details**

The sample structure used for the combined study of SPHE induced ultrafast demagnetization is shown in Fig. 1a. The sample is optimized so that the HE pulses are optically generated by ultrashort laser pulses using the Pt capping layer. The HE current pulses are then sent through the Cu (60nm) film, where the IR pulse is absorbed and through a hard-magnetic Co/Pt multilayer film, which generates ultrashort SPHE pulses[17,22]. The SPHE induced ultrafast demagnetization can then be probed in a soft-magnetic $Fe_{74}Gd_{26}$ layer (detector) located at the bottom of the spin valve. The sketch in Fig. 1b shows the orientation of the Fe and Gd sub-lattices magnetization ($M^{Fe}$ and $M^{Gd}$), coupled antiparallel in $Fe_{74}Gd_{26}$. The alloy has a compensation temperature above 350 K, which is significantly higher than the sample temperature before the IR pump excitation ($t < t_0$). In Fig.1b, the blue arrow indicates the Gd 4f magnetization, which can be aligned by an external magnetic field of moderate amplitude (H=100 Oe) well below the coercive field of the Co/Pt multilayer (550 Oe). The green arrow represents the Fe 3d magnetization, exchange coupled so that it is antiparallel to the Gd 4f magnetization. Note that P (AP) defines the parallel (antiparallel) orientations between the magnetization of the Co/Pt multilayer ($M^{CoPt}$) and the Fe sub-lattice magnetization ($M^{Fe}$) in $Fe_{74}Gd_{26}$ (Fig 1b).

The magnetic configuration at the thermodynamic equilibrium of the CoPt / Cu (10) / $Fe_{74}Gd_{26}$ spin-valve has been analyzed by X-ray magnetic circular dichroism (XMCD) measurements as a function of temperature, see Methods section. The 10 nm thick Cu film ensures no magnetic coupling between Co/Pt and $Fe_{74}Gd_{26}$ films. The spin valve is thus ideally suited to measure two relative spin-spin configurations by changing only the relative magnetizations in both magnetic films. Thanks to the chemical sensitivity of XMCD, our static measurements performed at the Gd $M_5$ and Co $L_3$ edges are used to define the sample temperature during the pump-probe experiments (for details, see Fig. 2 and 3 in Supplementary Information).

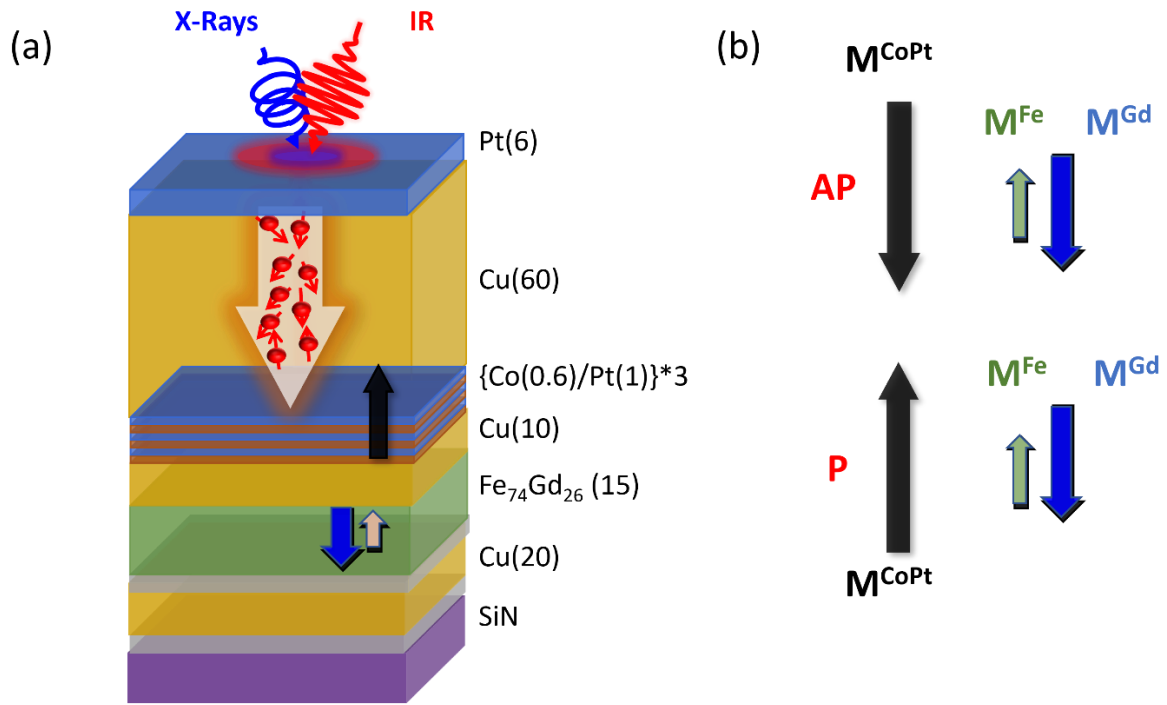

**Figure 1. Spin valve structure and schematic representation of experimental scheme: (a)** Sample structure used to study the spin-polarized hot electron (SPHE) induced dynamics: SiN/Ta (5) /Cu (20)/Ta (5) / Fe$_{74}$Gd$_{26}$ (15) /Cu (10) / {Co (0.6)/Pt (1)*3}/Cu (60) /Pt (6). The thickness of each layer in the bracket is in nm. Here, the red pulse represents the IR laser (800 nm) of 60 fs duration acting as a pump, and the blue pulse represents the circularly polarized X-ray pulses of 100 fs duration as a probe. Both pulses are separated by 1°. The large arrow represents the direction of hot electrons flow, and inside, the short red arrow with a circle represents the non-polarized hot electrons. Cu (60) ensures the complete absorption of the IR pulse; therefore, the bottom Co/Pt layer is excited through HE pulses only. Consequently, those HE pulses generate spin-polarized current (SPHE) from Co/Pt. After crossing the spacer layer, the spin current interacts with the Fe$_{74}$Gd$_{26}$ layer, on which the response of SPHEs is recorded. (b) Parallel (P) and antiparallel (AP) experimental scheme: The Black, green, and blue arrows represent the Co, Fe, and Gd magnetization directions. AP and P define the relative orientations of the magnetization between Co and Fe in {Co (0.6)/Pt (1)*3} and Fe$_{74}$Gd$_{26}$, respectively.

By measuring the hysteresis at Co L$_3$ and Fe L$_3$ edges (see inset of Fig.3a), we defined the saturation fields during pump-probe of both magnetic films (Co/Pt and Fe$_{74}$Gd$_{26}$) of 1000 Oe and 200 Oe, respectively. To distinguish the SPHE effect from overall thermal demagnetization, the pump probe experiment was performed in two experimental geometries, P and AP orientation of Co and Fe magnetization direction in {Co/Pt}*3 and Fe$_{74}$Gd$_{26}$ magnetic layer, respectively. The detailed experimental schemes to reach the P and AP geometries are explained in the method section. The X-ray transmission experiment, using an IR laser pump (red) and X-ray probe (blue) configuration, is schematically shown in Fig. 2. The external magnetic field is applied along the propagation direction of the X-rays (blue). The time $t_0 = 0$ is defined by the temporal overlap between the IR laser and X-ray pulses. The incident X-rays are circularly polarized. The TR-XMCD is extracted by making the difference of the transmitted X-ray absorption spectra intensities recorded by applying two opposite magnetic fields, +H and -H, parallel to the X-rays, as a function of time (see Methods). Alternating the pumped and the unpumped signals at the Fe L$_3$ and Gd M$_5$ edges allows for the normalization of the XMCD signal during the pump-probe delay scans. At the same time, it allows to verify the magnetization at negative delay in the Fe$_{74}$Gd$_{26}$ film.

**Ultrafast demagnetization in Fe$_{74}$Gd$_{26}$. The case of Fe.**

In Fig.3, we show the pump-probe results obtained at the Fe L$_3$ edge of the Fe$_{74}$Gd$_{26}$ alloy layer at an IR fluence of 120 mJ/cm$^2$. The values at negative delays are normalized to 1. The continuous lines are the results of the fitting (see Methods). The inset of Fig 3a shows the (static) element selective hysteresis measured at the Co L$_3$ and Fe L$_3$ resonance edges during the pump probe experiment at negative delays. This data indicates that the coercivity of the Co/Pt layer (H$_C$ = 550 Oe) is larger than that of FeGd (H$_C$ = 100 Oe). The measurements in Fig.3a show the ultrafast demagnetization dynamics induced by the SPHE pulses where magnetization of Co in Co/Pt layer is parallel (P) (blue curve) and antiparallel (AP) (red curve) to the magnetization of Fe in the Fe$_{74}$Gd$_{26}$ alloy. Solid lines are the exponential fit with the Gaussian convolution. We have analyzed the demagnetization dynamics obtained at the Fe L$_3$ edge at two timescales: At short time scale (Fig.3a), i.e., below 2.5 ps, we find that the experimental dynamics in the parallel case (P) is faster and slightly larger in amplitude than for the AP case. Characteristic demagnetization times are extracted for P and AP cases and have values of 470 ± 80 fs and 550 ± 80 fs, respectively.

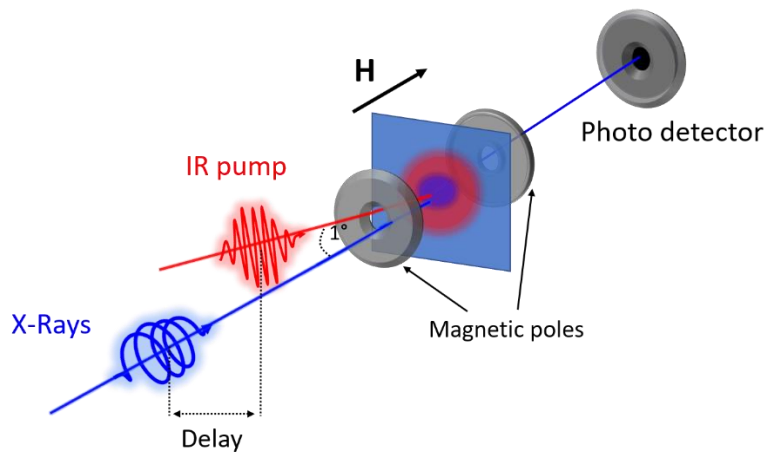

**Figure 2. Schematic illustration of tr-XMCD pump-probe experimental set-up:** Schematic of experimental set-up at Femtoslicing beamline, BESSY. IR laser (800 nm) of pulse width of 60 fs with a repetition rate of 3 kHz is used as a pump and ultrashort circular polarized X-rays of 100 fs duration and 6 kHz repetition rate is used as a probe. Both pump and probe are separated by 1° and incident normally on the sample. Here, X-ray absorption spectra are recorded in transmission geometry by an avalanche photodiode detector. A vector magnet (represented by a circular disk) is used to apply

an alternating magnetic field to measure the change in XMCD as a function of pump-probe delay. Here, a solid black arrow represents the direction of the applied field. Since, the magnetic film has out-of-plane anisotropy, therefore, applied field and X-rays are parallel to the easy axis of magnetization.

The observed change in the demagnetization dynamics between the P and AP cases, is related to the difference in the relative spin orientations between SPHE and the Fe sub-lattice magnetization in $Fe_{74}Gd_{26}$. The red Gaussian curve in Fig.3 represents the SPHE pulse shape (G with FWHM = 420 fs), arriving at the $Fe_{74}Gd_{26}$ layer. The observed different dynamics between P and AP last significantly longer than the presence of the pump pulse itself. In order to evidence the dynamics of the difference between P and AP configurations, we plotted in Figure 3b the normalized differences (open symbols and solid black line) using the experimental data as well as the fitted curves from Figure 3a. Note that the solid black line is calculated from the difference between the two fits shown in Fig 3a. These curves shown in Figure 3b are superposed with the smoothed normalized difference data (magenta dotted points) obtained by averaging over 15 adjacent points. This figure shows that the difference between P and AP dynamics starts as soon as the SPHE pulse excites the film at t = 0. The differences between P and AP configurations as shown in Fig.3b evidences that the maximum SPHE induced effect occurs at a delay of t = 0.5 ps. It is not clear if this maximum corresponds to the SPHE pulse shape or if SPHE induced effects develop specific ultrafast dynamics. By comparing with the pulse of the SPHE (G with FWHM = 420 fs), we highlight the fact that the spin-induced changes last up to t = 2 ps, much longer than the temporal superposition of the SPHE pump pulse. In Fig.3c, we show that at longer time scales, the difference in dynamics between P and AP orientations vanishes, which we assign to the limited mean free path and lifetime of the SPHEs. The dynamics of both orientations are therefore converged after ∼3 picoseconds, and the magnetic recovery proceeds in a similar way for both P and AP configurations. Fig.3 additionally, shows that the difference reduces after 1 ps, vanishing after 3 ps. These essential features will be analyzed below, using theoretical, atomistic spin dynamics simulations.

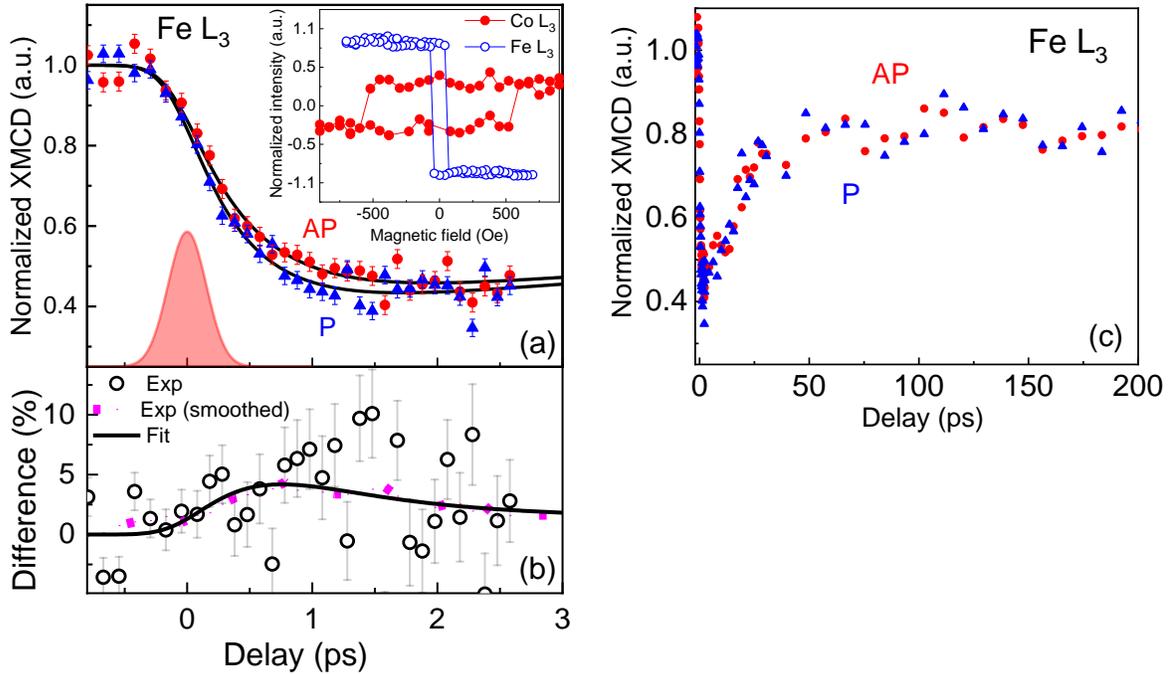

**Figure 3: Spin polarized hot electron induced dynamics at Fe $L_3$ edge**: Normalized time-resolved measurements of Fe $L_3$ XMCD at laser fluence of 120 mJ/cm$^2$ (absorbed fluence $F_{abs}$ = 3.2 mJ/cm$^2$) at T= 80 ± 20 K. To distinguish the SPHE effect, measurement is done in two configurations: Magnetization of Co is parallel (P) (blue curve) and Antiparallel (AP) (red curve) to Fe magnetization. Solid lines are the exponential fit with Gaussian convolution. The fitting of the P and AP experimental curves show characteristic demagnetization times of 470 ± 80 fs, resp. 550 ± 80 fs. (a) At a short time scale, i.e., below 2 ps, dynamics in the P case is faster and shows more demagnetization than AP case. The red Gaussian curve represents the SPHE pulse shape, arriving at the Fe$_{74}$Gd$_{26}$. The error bars obtained for the TR-XMCD at the Fe $L_3$ edge as shown in Figure 3a are given by the standard deviation of the experimental data with respect to the fitting functions. Inset: The hysteresis measured at the Co and Fe $L_3$ resonance edges as measured in the normal hybrid mode at the femtoslicing beam line. This indicates that the coercivity of Co/Pt layer ($H_C$ = 550 Oe) is larger than the FeGd ($H_C$ = 100 Oe) and ensures the maximum spin polarization from Co/Pt. (b) Normalized difference of the demagnetization dynamics between the P and AP cases obtained at the Fe $L_3$ edge (open symbols) superposed to the smoothed data obtained by averaging over adjacent 15 points (magenta dash line). The solid black line is the normalized difference between both fitted dynamics (P and AP exponential fitted curves) as shown in Fig 3a by solid black lines. A maximum for the difference between both cases is found around 0.5 ps. (c) Dynamics at a longer time scale. SPHE effect vanishes after 2 ps due to limited mean free path and lifetime of those SPHEs.

**Ultrafast demagnetization in Fe$_{74}$Gd$_{26}$. The case of Gd.**

In Fig.4, we show the same experiment with the focus on the Gd 4f moment, recorded by measuring the dynamics at Gd M$_5$ edge, at a fluence of 40 mJ/cm$^2$. The ultrafast dynamics measured in P (blue) and AP (red) configurations are shown for short (Fig.4a) and long (Fig.4b) time ranges. Note that the values for Fe (Fig.3) and Gd (Fig.4) are normalized to 1. The lower fluence used during the pump-probe experiments at Gd M$_5$ edge explains the fact that the amplitude of demagnetization is only about -35% compared to -55% at the Fe L$_3$ edge. This different demagnetization amplitude is not related to other physical or chemical reasons (see S.I. - Fig. 4, comparison of Fe – Gd dynamics). Within the achieved experimental noise level Fig.4 evidences no difference between P and AP configurations. We thus show only a double exponential fit with a characteristic demagnetization time of 900 ± 50 fs, typical times for FeGd alloys. Theoretical calculations (see below) also show that the expected differences between P and AP should be smaller for Gd M$_5$ than for Fe L$_3$, when using fluences that are appropriate to the experimental data of Fe and Gd. The larger X-ray cross section at the Gd M$_5$ edge compensates for the lower concentration in Gd M$_5$ compared to Fe L$_3$ and cannot explain the larger noise level at Gd M$_5$. However, this noise level can be attributed to the lower experimental laser fluence and, thus, to lower demagnetization amplitude compared to for Fe L$_3$. Pumping at substantially higher fluences resulted in experimental conditions close to the destructive limit. Data of high statistical quality for this time- and sample-consuming high-fluence operation could only be obtained for the Fe L$_3$ edge. We note that the slicing facility's X-ray pulse stability can vary from week to week, contributing to different noise levels for similar acquisition statistics.

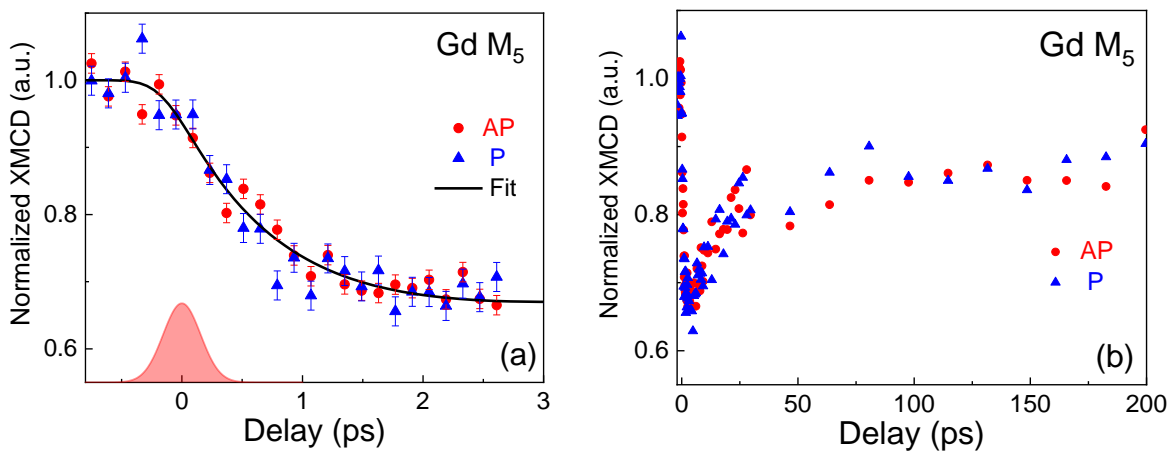

**Figure 4. Spin polarized hot electron induced dynamics at the Gd $M_5$ edge, for short (a) and long (b) time scales**. (a) Normalized experimental time-resolved measurements of Gd $M_5$ XMCD were obtained from the transmission signal at a laser fluence of 40 mJ/cm$^2$ (absorbed fluence $F_{abs}$ = 1.2 mJ/cm$^2$) and at $T$ = 140 K working temperature. The measurements are done in two configurations: Magnetization of Co is parallel (P) (blue curve) and Antiparallel (AP) (red curve) to Fe magnetization. The estimated error bars are the standard deviation of the experimental data with respect to the fitting function. The solid black line represents the exponential fit with Gaussian convolution. We obtained the characteristic demagnetization time, $\tau_{Gd}$ = 900 ± 100 fs. (b) Dynamics at a longer time scales showing similar dynamics for P and AP. Note that the statistics is not the same between 30 ps and 200 ps than at short time scales.

**Simulated ultrafast demagnetization in $Fe_{74}Gd_{26}$.**

To study the ultrafast demagnetization dynamics of $Fe_{74}Gd_{26}$, we performed atomistic spin-dynamics simulations[40] (see Methods section for simulations details). To analyze the experimental data, we have created a simulation cell consisting of 1600 atoms distributed inhomogeneously. In these simulations, we have followed theoretical and experimental studies suggesting that in the amorphous sample investigated here the concentration of Fe and Gd varies[41,42]. In particular, we consider an amorphous alloy with an average concentration of $Fe_{74}Gd_{26}$ (as shown in Supplementary Information), this gives a pair distribution function that matches experimental data (see Ref.42 for details). However, these alloys have a modulation of the chemical composition and to accommodate this we considered chemical modulations so that some areas are 6% richer in iron and some are 6% richer in gadolinium than the nominal concentration of $Fe_{74}Gd_{26}$. In our simulations, we assume that the hot electron pulse leads to an increased electron temperature in the $Fe_{74}Gd_{26}$ sample. The electronic temperature rise is accounted for by using a three-temperature model (3TM), that allows for heat to flow between electron-, lattice- and spin reservoirs. The 3TM was proposed in the pioneering work of Beaurepaire et al.[6] to calculate spin, lattice, and electron temperatures during ultrafast demagnetization dynamics. The model assumes three thermalized reservoirs, in particular, spin, lattice, and electron connected by electron-spin $G_{es}$, electron-lattice $G_{el}$, and spin-lattice $G_{sl}$ coupling coefficients. When the electronic temperature increases (in the present case due to the hot electron pulse), the rise of spin- and lattice temperatures is mediated by these coupling coefficients $G_{es}$, $G_{el}$, $G_{sl}$. Parameters and details can be found in the Methods section and Supplementary Information. The temperatures calculated using 3TM are then used in atomistic spin dynamics simulations (see Method Section). To analyze as closely as possible the experimental measurements presented above, we study separately iron and gadolinium magnetization dynamics in amorphous $Fe_{74}Gd_{26}$. The exchange interactions used in the simulations are reported in Ref.42, and were selected to reproduce the static ordering temperature as well as the measured compensation temperature. The other parameters for our simulations are to some degree chosen to accomplish a good comparison with the experimental data by carefully studying the impact of simulation parameters on the magnetization curves. In particular, the spin transfer torque, Gilbert damping $\alpha$, electron-spin-, electron-phonon, and spin-lattice coupling in the three temperature models, will impact all the resulting aspects of the magnetization dynamics. The details of how these parameters influence the dynamics of $Fe_{74}Gd_{26}$ are shown in Supplementary Information.

After optimization of the simulated dynamics with our experimental results, we obtained, with one marked difference that we will return to below, a reasonable agreement between simulations and the experimental data recorded at the Fe $L_3$ edge during the first ~2 picoseconds of the magnetization dynamics (Fig.5). The sensitivity of the simulated data with respect to parameter choice is analyzed in detail, shown in Figs. 6-11 in Supplementary Information. Most notable the simulations shown in Fig.5 reproduce the overall shape of the experimental curves, both when it comes to the position of the minimum of the $M^{Fe} / M^{Fe}_0$ curve (at 1.5 - 2 ps) as well as the overall shape of the $M^{Fe} / M^{Fe}_0$ demagnetization curve. In fact, simulations and experiment are basically on top of one another, with one marked exception, the simulated results with parallel STT fit perfectly the experimental results for AP configuration and simulated results with antiparallel STT fit perfectly the experimental results for P configuration, which is contrary to what is expected from the angular momentum transfers schematically shown in Fig.1. We will return to this enigma below.

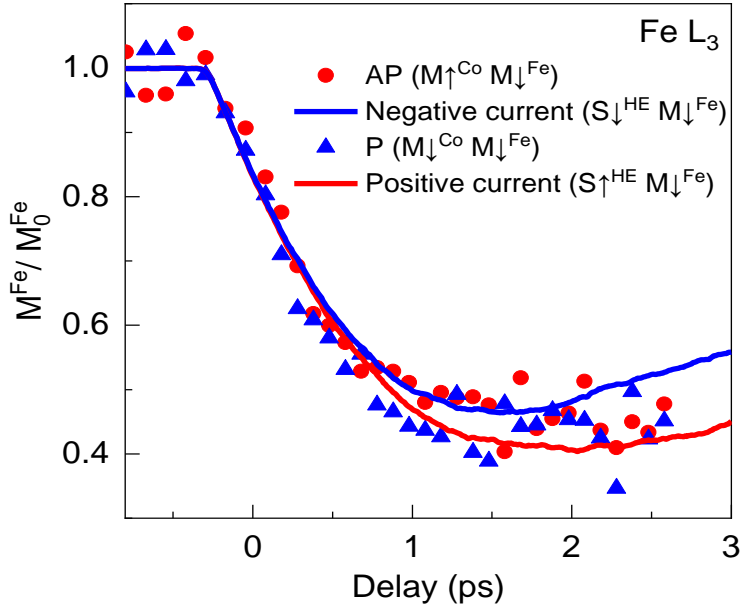

**Figure 5. Magnetization dynamics of Fe in amorphous $Fe_{74}Gd_{26}$.** Experimental and Simulated demagnetization dynamics by TR-XMCD and atomistic spin dynamics of Fe in $Fe_{74}Gd_{26}$ for 120 mJ/ cm² incident and 3.2 mJ/cm² absorbed fluences. The laser fluence value has been adjusted to fit our experimental data. The blue (red) symbols are the experimental results for the P(AP) configurations. Solid lines in red (blue) are the simulations for the antiparallel (parallel) STT and the associated positive (negative) spin currents. In simulations the spin current polarization ($S^{HE}$) is defined as "positive" for the antiparallel orientation to the Fe 3d magnetization of the Fe sub-lattice ($M^{Fe}$) so that the SPHE polarization ($S^{HE}$) is opposite to the Co magnetization ($M^{Co}$), shown by the arrows in the figure legends.

Similarly, the simulated results for gadolinium are also in very good agreement with experimental measurement, as shown in Fig.6. In the simulations for Gd, we adjusted the heat-driven dynamics induced by the absorbed HE pulse by a factor of 0.4 in comparison to the data for Fe, which corresponds to the reduction of the IR incidence fluence between both experiments (The nominal fluence ratio is somewhat lower: 40 mJ / 120 mJ = 0.3). We assign the difference to the limited accuracy of laser fluence determination ±20%, mostly because of uncertainties with the laser spot-size. Parameters in the simulations, such as Gilbert damping, and heat transfer parameters of the 3TM, influence the demagnetization amplitude of both sub-lattices. These can in the experimental samples differ for iron and gadolinium, while in the simulations presented here, we use for simplicity the same values for both sub-lattices. This may lead to a slight underestimation/overestimation of Fe and Gd demagnetization amplitudes. By deviating from the experimental estimate of the heat provided by the hot electrons, we compensate for these differences. These results in simulated data which are in rather good agreement with observations (see Fig.6).

We now focus on the role of Gd in the amorphous $Fe_{74}Gd_{26}$ alloy. As shown in Figure 6, the demagnetization for gadolinium shows smaller amplitudes and a slower demagnetization dynamic than what is observed for Fe (Fig.5) while the experimental statistics does not allow drawing conclusions about the difference between P and AP cases. However, from the simulations, it can be evidenced that for gadolinium, the parallel and antiparallel STT simulations lead to slightly different demagnetization dynamics (Fig.6). Even with the small difference in the simulated curves, we are able to demonstrate that the acceleration of the demagnetization appears for a STT opposite compared to that for Fe, which is consistent with the ferrimagnetic state of the FeGd alloy investigated here.

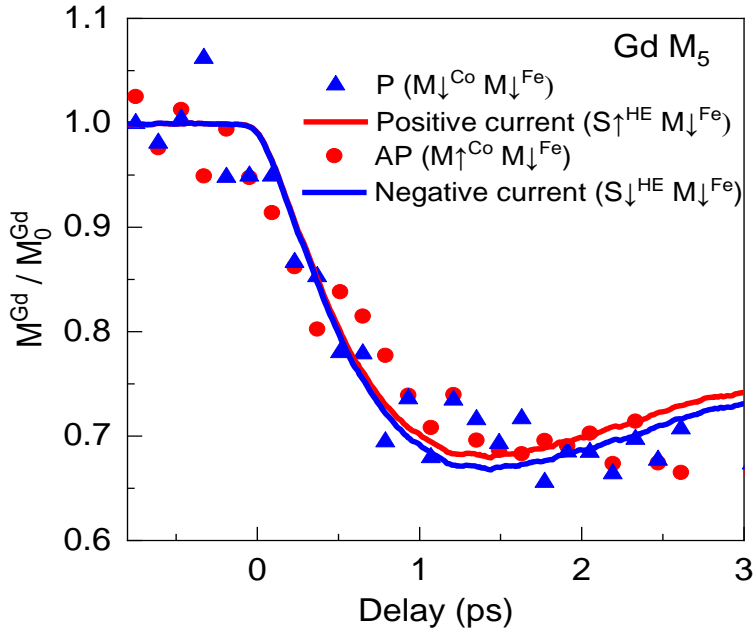

**Figure 6. Magnetization dynamics of gadolinium in amorphous $Fe_{74}Gd_{26}$.** Experimental and Simulated demagnetization dynamics by TR-XMCD and atomistic spin dynamics of Gd in $Fe_{74}Gd_{26}$ for 40 mJ/ cm² incident and 1.2 mJ/cm² absorbed fluences. The last fluence has been adjusted to fit our experimental data. The blue (red) symbols and lines are the results for the P (AP) configurations. Solid lines in red (blue) are the simulations for the antiparallel (parallel) STT and the associated positive (negative) spin currents. In simulations the STT spin current polarization ($S^{HE}$)
15

is defined as "positive" for the antiparallel orientation to the Fe 3d magnetization of the Fe sub-lattice ($M^{Fe}$).

In order to investigate if the absence of splitting between curves labelled P and AP observed at the Gd $M_5$ edge is only due to the lower fluence used for Gd, the theoretical simulations were extended to absorbed fluences of 3.2 mJ/cm$^2$, which is same as used in the simulations for Fe $L_3$ (Fig.5). Figure 7 shows the so-simulated dynamics for Gd 4f moments for parallel and antiparallel STT. The limited, but discernible, differences observed for Gd at these larger fluences indicate that significantly better experimental statistics would be needed at the Gd $M_5$ edge to evidence these small effects. In the experimental results reported here, taken at the slicing station at BESSYII, this was not possible to achieve.

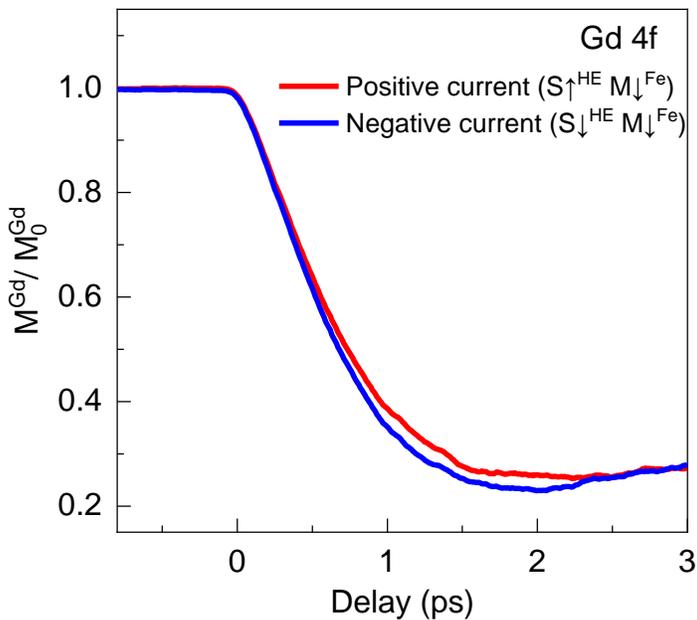

**Figure 7: Calculated magnetization dynamics of Gd 4f in amorphous Fe$_{74}$Gd$_{26}$.** Compared STT effect (difference between positive and negative STT) for both absorbed fluences at Gd in amorphous Fe$_{74}$Gd$_{26}$ for 120 mJ/cm$^2$ incident and 3.2 mJ/cm$^2$ absorbed fluences. The red and blue lines are simulations for opposite signs of spin polarized currents (positive and negative) combining heat-driven demagnetization dynamics with STT. Gilbert damping value α = 0.1.

Let us now extend the analysis to the P/AP effect in the FeGd alloy. Using the calculated demagnetization curves obtained for Fe and for Gd at the absorbed fluence of 3.2 mJ/cm$^2$, we can extract the dynamics for P and AP in the case of the Fe$_{74}$Gd$_{26}$ alloy (Fig 14 in S.I.). We use a linear combination of the P/AP effect for Fe and Gd including the concentration and atomic magnetic moments in each sublattices. Here we use the same magnetic moment values as used in our simulations[42]. These simulations highlight that the ferrimagnetic alloy leads to a large P/AP effect (normalized difference of ~38 %). The large effect is explained by the ferrimagnetic order of Fe and Gd, strongly reducing the total magnetic moment, combined with the opposite P/AP effect in each of the sublattices. We note that in the simulations, at t= ~ 0.4 ps, the overall demagnetization M$^{FeGd}$ changes its sign and orients parallel to the Fe sub-lattice magnetization. In these simulations the change of sign is related to the crossing of the compensation temperature, where the Fe and Gd atomic moments are equal but opposite in sign. The switching of the total moment in the alloy below 1ps depends essentially on the concentration of the alloy and is not affected by small changes in the atomic magnetic moments of Gd or Fe. Experimentally, we could not observe this effect because we did not measure the Gd M edge at high fluence.

## Discussion

In this work, we have investigated experimentally the induced effect of ultrashort pulses of SPHE on the ultrafast demagnetization of amorphous Fe$_{74}$Gd$_{26}$ with an out-of-plane magnetic anisotropy. The pump-probe measurements at Fe L$_3$ edge reveal not only faster demagnetization dynamics of the Fe 3d moments in the case of P configuration compared to AP configuration but also a reduction in the demagnetization amplitude for the AP configuration. A similar experiment performed at Gd M$_5$ edge did not show a sizable impact of spin polarized electrons, most probably due to weaker P/AP effects. We would like to stress that the experimental findings as such are unique since the employed time-resolved X-ray spectroscopy distinguished the induced effect of ultrashort pulses of SPHE on rare-earth Gd from the ones on transition metals Fe. The optical IR induced ultrafast dynamics have been studied already in the past in RE-TM alloys[2,29,41,43,44] but as far as SPHE induced effects are concerned, the induced dynamics and differences between RE and TM elements have never been addressed.

Accompanying experiments, we employed atomistic spin dynamics simulations that utilized the Landau-Lifshitz Gilbert (LLG) equation. In these simulations it was assumed that the hot electron pulse leads to a rise of the electronic temperature of $Fe_{74}Gd_{26}$, followed by an increase in lattice- and spin temperature that we model with the three-temperature model (3TM). In addition, the atomistic modeling considered a dynamic mechanism based on the spin transfer torque (STT), where the spin angular momentum of an electric current couples with local magnetic moments of the material. Within this framework, it was found to be possible to reproduce the acceleration of the dynamics observed for the Fe sublattice and predict the effects for the Gd sublattice. The calculations show that, using a limited set of reasonable parameters (see Table II in Supplementary Information) belonging to the ferrimagnetic $Fe_{74}Gd_{26}$ alloy, the time scales and amplitudes of spin dependent demagnetization can, with one noticeable exception to be discussed shortly, be reproduced. As the spin-transfer torque prefactor used in the simulations includes both spin current density and polarization, we simulate the demagnetization dynamics (Fig.10 of S. I.) by either changing the current density or polarization to determine the range of these physical parameters. It can be seen from Fig. 10 (S.I.) that changes around ± 25 % in current density or polarization do not impact the observed difference between the P and AP configurations and preserve the conclusions of the manuscript. The simulations show a very important result considering the spin polarization induced demagnetization dynamics in Fe and Gd sublattices. A STT with a spin polarization of the SPHE parallel to the Co magnetization (Fig.1) results in opposite dynamics to that observed in experiments, when it comes to the splitting between dynamics of the P and AP configurations (Fig.5). In contrast, a STT with a spin polarization coupled antiparallel to the Co magnetization reproduces almost perfectly the dynamics of the here investigated system, including the splitting between P and AP couplings. These results show similar trends as observed by Igarashi et al., in a spin valve structure[19]. Based on our work and theoretical calculations, we conclude that the outcoming SPHE current from the Co/Pt stack has opposite spin polarization than the Co magnetization and that this explains the experimental results. The microscopic mechanism behind the opposite Co magnetization and the spin polarization of the electric current (used in STT) can be traced to the minority spin polarization of electron states close to the Fermi level of fcc Co[45]. The mechanism we hence propose is that SPHE generated by the 1.5eV laser pump, thermalizes during the propagation of top Pt(6)/ Cu(60) layers. We know that only during the propagation through Pt the thermalization can happen whereas in Cu much less scattering of the HE should happen. The multiple scattering during the propagation through 6 nm Pt could thus lead to the thermalization (E ~0.1 eV) of the HE current towards the Co/Pt polarizer, where minority spin-polarization is generated.

Finally, considering simple macroscopic arguments of angular momentum conservations in the spin valve, we estimate that a maximum of angular momentum loss (~ 4.5 $\mu_B$) in Co/Pt will be taken out from the multilayer (3nm thick Co film times 1.5 $\mu_B$/at) which can be transferred to the FeGd alloy (at t = $t_0$, a 15nm thick FeGd film, times 2 $\mu_B$/at = 30 $\mu_B$). We thus expect at maximum a change of 12% in the angular moment of the FeGd alloy due to the spin effect in HE induced demagnetization, assuming a complete quenching of the Co/Pt film and a 100 % transfer efficiency.

Qualitative estimations of the relative effects in Fe and Gd in the FeGd alloy can be discussed from a simple energy point of view. In the simulations we propose a double contribution to the demagnetization, where the first is heat-driven by the HE energy transfer to the electronic system, whereas the second is provided by the energy transferred by the spin polarization of SPHE, and is modeled in our work by a STT. Double pulse induced dynamics has also been reported recently to explain switching mechanisms[19,46,47]. Considering previous published results[43], we expect that in RE-TM alloys the RE and TM sublattices show opposite changes (acceleration or deceleration of the dynamic) when temperature is increased towards $T_C$[43,48,49]. In $Fe_{74}Gd_{26}$, we have $T_C$ = 500K which leads to: Tc-T = 500 - 140 = 350 K. The proximity to $T_C$ could thus explain similar characteristic times for Fe and Gd sublattices as shown in S.I. figure 4 for HE induced dynamics. In FeGd assuming that in P configuration (blue symbols in Fig. 5), a supplement of energy flows into the 3d Fe, compared to AP, then we could expect from a temperature dependent model, that the excess energy raises the electron temperatures in the FeGd alloy leading to an acceleration of 3d Fe and to a deceleration of 4f Gd dynamic. This is what is observed here for FeGd. However, this is opposite to the previously studied CoDy system which demonstrates that the temperature dependent accelerations - decelerations are compositionally dependent on the RE-TM system[43].

**Methods**

Sample preparation and magnetic properties of the alloy films: 15nm thick alloys have been grown by magnetron sputtering on $Si_3N_4$ membranes. Co-deposition with convergent Co, Pt and Fe, Gd flux was used to get amorphous $Fe_{74}Gd_{26}$. alloy films and Co/Pt multilayers.

Static XMCD and magnetic hysteresis were performed at the Co $L_3$, Fe $L_3$, and Gd $M_5$ edges at the ALICE Station at the PM3 beam line of the BESSY II synchrotron radiation source of the Helmholtz-Zentrum Berlin[50]. Time-resolved XMCD: Time resolved XMCD was performed at the femtoslicing beam line of the BESSY II synchrotron radiation source of the Helmholtz-Zentrum Berlin[7,33]. Two different operation modes can be used at the beam line: the "normal" mode (50 ps time resolution) with large X-ray flux to monitor static magnetic properties (hysteresis) and the femtoslicing operation mode with $10^3$ times less flux with 130 fs total time resolution. The magnetization dynamics have been measured by monitoring the transmission signal of circularly polarized X-rays, tuned to specific core level absorption edges as a function of a pump-probe delay using femtoslicing mode. The dynamic XMCD contrast is obtained by subtracting the gated signals obtained with and without pump beam. The energy was set to the different Fe $L_3$, and Gd $M_5$ edges using the Bragg Fresnel reflection zone plate monochromator. The experiments have been performed with a pump-probe setup where the short X-ray pulses are synchronized with a femtosecond pump laser working at 800 nm, 3 kHz repetition rate with pulses of 60 fs. The X-ray pulse duration of about 100 fs in the femtoslicing operation mode ensures a global time resolution of ~ 130 fs (see refs.[7,33] for details). The pump fluences used during our experiments were adjusted to 120 $mJ/cm^2$ for the study of the dynamics at the Fe $L_3$ edge and to 40 $mJ/cm^2$ for the Gd $M_5$ edge to get the demagnetization magnitudes of about 50% at the Fe $L_3$ edge and 35% at the Gd $M_5$ edge without altering the sample properties (alloy concentration, atomic diffusion, large DC heating).

Detailed field sequences used for P and AP experimental schemes: Since the coercive field of Co/Pt layer ($H_C$ = 550 Oe) is 5 times larger than the soft magnetic FeGd layer ($H_C$ = 100 Oe), the bilayer acts as a spin-valve. The distinct Hc values allow to switch the FeGd layer without affecting the magnetization state in the hard-magnetic layer Co/Pt, thus aligning the magnetization in both layers independently.

1. To measure the TR-XMCD in the AP configuration, a magnetic field H = +1000 Oe is applied and the transient XMCD is recorded by alternatively recording the transmission signal at +H and -H at Fe L3 and Gd M5 edges.

2. The P configuration is obtained by applying $H_1$ = +1000 Oe, followed by an opposite smaller field $H_2$= - 200 Oe. This aligns the Gd magnetization in the opposite direction without changing the initial [Co/Pt]*3 magnetization. As a consequence, the Fe magnetization aligns parallel to Co magnetization (defined as the P case). We measured the transient XMCD= (XAS+ - XAS-) at the FeL3 and GdM5 edges in the AP state by a series of two successive field pulses of -1000 Oe followed by +200 Oe and +1000 Oe followed by- 200 Oe so that we ensure the full saturation in both magnetic films [Co/Pt]*3 and FeGd.

**Fitting procedure:**

The physical quantities M(t) were derived (Fig. 3 and 4), using the rate equation of the two-temperature model with two exponential functions (equation 1):

$$F(t) = G(t)*(C_0 + C_1 H(t-t_0)[1 - exp(-(t-t_0)/\tau_{th})]exp(-(t-t_0)/\tau_s) \qquad (1)$$

where G(t) is the Gaussian function defining the total time resolution of the experiment (425 fs), $\tau_{th}$ and $\tau_{s-ph}$ are the thermalization time and the relaxation time from the spin system to other systems (lattice, external bath), $t_0$ is the delay at which the temporal overlap of the pump and the probe is achieved and $H(t-t_0)$ is the Heaviside function ($H(t-t_0) = 0$ if $t < t_0$, and $H(t-t_0) = 1$ if $t > t_0$) describing the energy transfer from the laser.

**Atomistic spin dynamics simulations**

In atomistic spin dynamics simulations, spin dynamics is governed by Landau-Lifshitz-Gilbert equation:

$$\frac{d\boldsymbol{m}_i}{dt} = -\frac{\gamma}{1+\alpha^2}\boldsymbol{m}_i \times \left(\boldsymbol{B}_i + \boldsymbol{B}_i^{Fl}\right) - \frac{\gamma}{(1+\alpha^2)m_i}\alpha \boldsymbol{m}_i \times \left(\boldsymbol{m}_i \times \left(\boldsymbol{B}_i + \boldsymbol{B}_i^{Fl}\right)\right) \qquad (2)$$

where $\boldsymbol{m}_i$ represents an atomic magnetic moment, $m_i$ and $\gamma$ are the magnitude of the atomic magnetic moment, and the gyromagnetic ratio correspondingly. $\alpha$ is the Gilbert damping parameter. We obtain an effective exchange field $B_i = -\partial H_{SD}/\partial m_i$ from the spin Hamiltonian, $H_{SD}$. In our simulations, we use a stochastic field, $B_i^{Fl}$, as white noise with properties

$$\langle B_{i,\mu}^{Fl}(t) B_{j,\nu}^{Fl}(t') \rangle = 2D_i \Delta t \delta_{ij} \delta_{\mu\nu} \delta(t-t')$$

Here, i and j denote lattice sites, μ and ν are the Cartesian components, noise power $D_i = \alpha k_B T_e/(1+\alpha^2)\gamma m_i$, where $T_e$ and $k_B$ are electronic temperature calculated from 3TM model (explained below) and Boltzmann constant respectively (please see Ref.[40]), $\Delta t$ is a time step. The formalism above is implemented in the UppASD[51] code which was used for all simulations in this work.

In our simulations we use the Heisenberg model with the magnetic Hamiltonian described by:

$$H_{SD} = - \sum_{<i,j>} J_{ij} \mathbf{m}_i \cdot \mathbf{m}_j \tag{3}$$

where $J_{ij}$ is the exchange interaction between magnetic moments $m_i$ and $m_j$.

We couple the atomistic dynamics to the three-temperature model[6], proposed by Beaurepaire, to describe the electronic heat bath that drives our Langevin-based ASD simulations[14,37,40]. The model assumes three reservoirs: electronic, spin, and lattice with corresponding temperatures $T_e$, $T_s$, and $T_l$, these three reservoirs can exchange heat via heat transfer coefficients spin-lattice $G_{sl}$, electron-spin $G_{es}$, and electron-lattice $G_{el}$ (please see Supplementary Information Table II for the values used in calculations) the temperature dynamics of the system are then governed by the three coupled differential equations:

$$C_e(T_e)\frac{dT_e}{dt} = -G_{el}(T_e - T_l) - G_{es}(T_e - T_s) + P(t) \tag{4}$$

$$C_s(T_s)\frac{dT_s}{dt} = -G_{es}(T_s - T_e) - G_{sl}(T_s - T_l) \tag{5}$$

$$C_l(T_l)\frac{dT_l}{dt} = -G_{el}(T_l - T_e) - G_{sl}(T_l - T_s) \tag{6}$$

Where $C_e$, $C_s$, and $C_l$ are the heat capacities of three subsystems, and P(t) represents the heat source, in our simulations, a hot electron pulse increases the electronic temperature. The resulting electron temperature $T_e$ calculated using 3TM is then used as the heat-bath for the atomistic spin dynamics simulations of ultrafast magnetization dynamics. In this temperature model, the spin- and lattice temperatures thus only enter as auxiliary variables in order to obtain a realistic model of the heat-bath which is determined by the electron temperature.

Spin-transfer torque (STT) was proposed by Slonczewski and Berger for a description of the impact of incoming itinerant electrons on localized magnetic moments in magnetic materials (see ref.[52] and references therein). STT is taken into account by adding the following field to the Landau-Lifshitz-Gilbert equation:

$$\mathbf{B}_i^{STT} = B^{STT}(\mathbf{m}_i \times \hat{\mathbf{p}})  \tag{7}$$

where the strength of the STT term $B^{STT}$ depends on the current density $j_e$ (Am$^{-2}$), and the spin polarization p while $\hat{p}$ is the unit vector pointing along the polarization axis. We considered the largest term from Eq. 12 in[52] since the other terms are found to be negligible for some layered structures[52]. The values of $j_e$ used in the simulations can be found in Supplementary Information.

## Data availability

Correspondence and requests for materials should be addressed to C. Boeglin Address: IPCMS, 23, rue du LŒSS, F-67034 STRASBOURG Cedex 02, France e-mail: christine.boeglin@ipcms.unistra.fr

## Acknowledgments

We are indebted to R. Mitzner and M. Mawass for help and support during the femtoslicing experiments. This work was supported by funding from the European Union's Horizon 2020 research and innovation programme under the Marie Skłodowska-Curie grant agreement number 847471, by the "Agence Nationale de la Recherche" in France via the project ANR-20-CE42-0012-01 (MEDYNA), ANR-21-CE42-0004-01 (EXPERTISE) and by the German Ministry of Education and Research BMBF Grant 05K10PG2 (FEMTOSPEX). This work was financially supported by Olle Engkvist Foundation. Support from STandUP and eSSENCE is also acknowledged. O.E. also acknowledge support from the Swedish Research Council, the European Research Council (ERC) as well as the Wallenberg Initiative Materials Science for Sustainability (WISE) funded by the Knut and Alice Wallenberg Foundation (KAW), the Swedish Research Council and the Knut and Alice Wallenberg Foundation (KAW). Computations were enabled by resources provided by the Swedish National Infrastructure for Computing (SNIC/NAISS) at NSC, partially funded by the Swedish Research Council through grant agreement no. 2018-05973. R.A. acknowledge financial support by the German Federal Ministry of Education and Research BMFG project number 05K19WO61.

## Author contributions statement

D.G., M.R., N.P., C.S-L., N.B. and C.B. performed the time resolved measurements and data exploitations. M.H. grew and characterized the samples. D.G., N.B., R.A. and C.B. performed the static temperature XMCD characterization. D.G. and C.B. wrote the first draft of the paper. A.B., O.E., and M.Pe initially designed the theoretical model. M.Pa. performed the simulations and calculations, M.Pa, A.B., B.S., S.E., O.E., and M.Pe analyzed the results and their interpretations. All authors discussed and contributed to the manuscript.

# Tuning of the ultrafast demagnetization by ultrashort spin polarized currents in multi-sublattice ferrimagnets.

**Deeksha Gupta**[1], **Maryna Pankratova**[2,6], **Matthias Riepp**[1], **Manuel Pereiro**[2], **Biplab Sanyal**[2], **Soheil Ershadrad**[2], **Michel Hehn**[3], **Niko Pontius**[4], **Christian Schüßler-Langeheine**[4], **Radu Abrudan**[4], **Nicolas Bergeard**[1], **Anders Bergman**[2], **Olle Eriksson**[2,5], **and Christine Boeglin**[1*]

[1]Institut de Physique et de Chimie des Matériaux de Strasbourg, UMR7504, CNRS et Université de Strasbourg, 67034 Strasbourg, France

[2]Department of Physics and Astronomy, Uppsala University, Box 516, SE-75120 Uppsala, Sweden

[3]Institut Jean Lamour, Université Henri Poincaré, Nancy, France

[4]Helmholtz-Zentrum Berlin für Materialien und Energie GmbH, Albert-Einstein Str. 15, 12489 Berlin, Germany

[5]Wallenberg Initiative Materials Science for Sustainability (WISE), Uppsala University, Box 516, SE-75120 Uppsala, Sweden

[6]Department of Engineering Sciences, University of Skövde, SE-541 28 Skövde, Sweden

[*]christine.boeglin@ipcms.unistra.fr

# I. EXPERIMENTAL DETAILS

## A. Calculation of absorption profile

We used the transfer matrix method [1] to calculate the absorption profile of the IR pulse in the sample structure (shown in Figure 1).

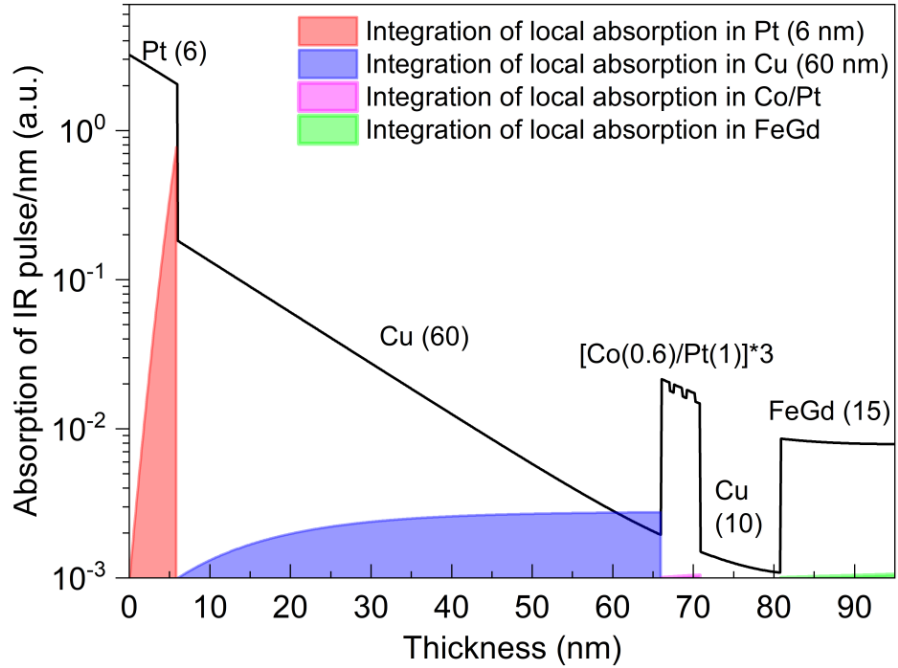

Figure 1. Absorption of IR intensity through the spin-valve structure. It shows that less than $10^{-3}$ is transmitted to the FeGd layer.

| Elements | Refractive Index (800 nm) |
|---|---|
| Pt | $2.83 + 4.95\,i$ [3] |
| Cu | $0.26 + 5.26\,i$ [3] |
| Co | $2.53 + 4.88\,i$ [2] |
| Fe | $2.94 + 3.39\,i$ [2] |
| Ta | $1.09 + 3.73\,i$ [3] |

Table I Refractive indices at 800 nm wavelength used to calculate the absorption profile.

## B. Variation of XMCD and coercive field with temperature

To set the experimental parameters, we performed static XMCD measurements using the ALICE reflectometer situated at the PM3 beamline at the BESSY II synchrotron source[4]. XAS spectra were recorded by employing two opposite magnetic fields (±H) in transmission geometry at Fe $L_3$, Co $L_3$ and Gd $M_5$ edges (fig 2 a). Additionally, we recorded the hysteresis of each element at their corresponding resonance edges as a function of temperature (fig 2 b). It is shown from the coercive field values in figure 2 that compensation temperature is above 300 K.

### C. Estimation of sample temperature during pump-probe experiment

Time-resolved XMCD measurements were performed at the cryostat temperature of 80 K. Due to limited photon flux during slicing mode, it's not possible to measure the hysteresis as a function of pump-probe delay. Therefore, in order to examine the sample temperature during the pump-probe experiment, we measured the time-resolved hysteresis using X-ray pulses of duration ~50 ps in the normal synchrotron mode, before switching to the femto-slicing mode. Figure 3 shows the hysteresis measured with Laser off and Laser on at t = 50 picoseconds (ps). By comparing the coercive field values from the static measurement data (fig 2), the temperature at the delay, t = 50 ps, is calibrated to be 220 K (represented by the cyan color dot in Fig 2).

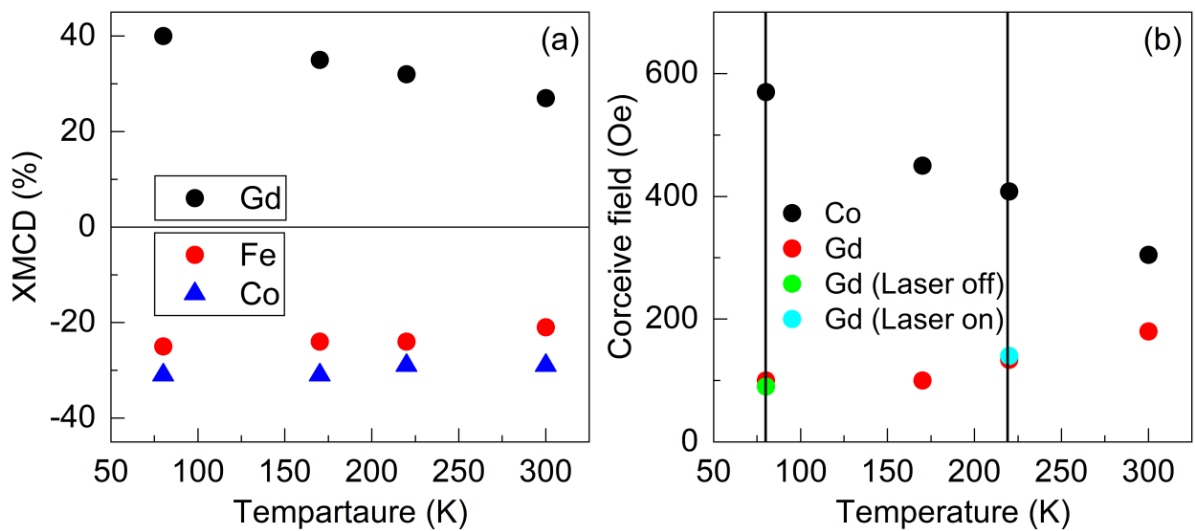

Figure 2. (a) Shows static XMCD (black (Gd M5), red (Fe L3), and blue (Co L3) points) as a function of temperature (b) shows coercive fields (black and red points) as a function of temperature for the spin-valve structure at the Co L3 and Gd M5 edges. As supplement information, we show (green and cyan

points) the coercive fields measured at the GdM5 edge, comparing the coercive fields without laser and with the laser (at a delay of t = 50ps). We observe that the coercive field at Gd M5 using the laser matches the static coercive field at T= 220K. This data defines the sample temperature during pump-probe experiments at a delay of 50 ps after the IR and HE excitation. Note that T= 80 K is the cryostat temperature without any additional DC heating, which may occur during the pump-probe experiment.

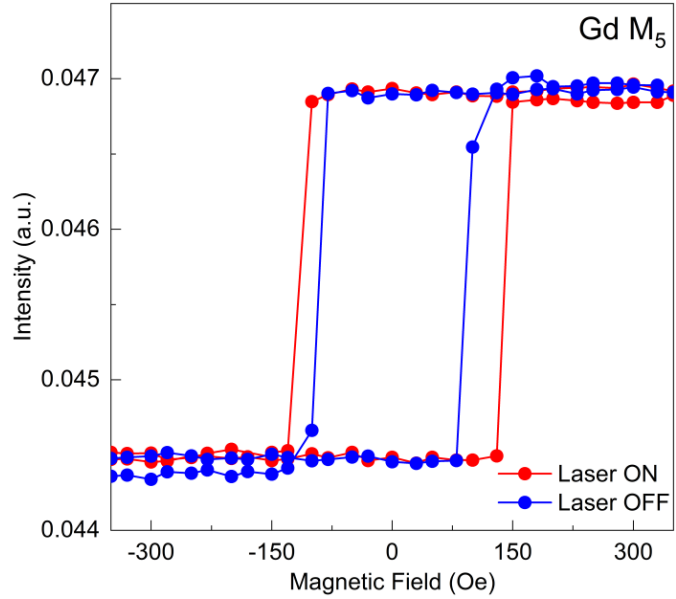

Figure 3. Hysteresis recorded at the Gd $M_5$ edge with and without the laser at a delay of t = +50 ps at an incident IR fluence of 40 mJ /cm$^2$. The sample temperature is set to 80K, below the magnetic compensation temperature ($T_M$) and we measure a coercive field Hc slightly below 100 Oe (blue curve). during pump-probe measurement the temperature increase towards $T_M$ and the coercive field of the ferrimagnetic layer rises slightly above 100 Oe [ 5].

Since, $Fe_{74}Gd_{26}$ alloy has a magnetic compensation temperature $T_M$ close to 350 K [REF 5] on increasing temperature and approaching $T_M$, the coercive field of the magnetic layer increases. As we mentioned in the S.I. -Fig3, during pump-probe measurement the temperature increases and therefore Hc becomes larger.

### D. Magnetization dynamics of $Fe_{74}Gd_{26}$

Figure 4 shows the magnetization dynamics measured at IR laser fluence of 40 mJ/cm$^2$. These measurements were performed during the same beamtime and with the same experimental conditions.

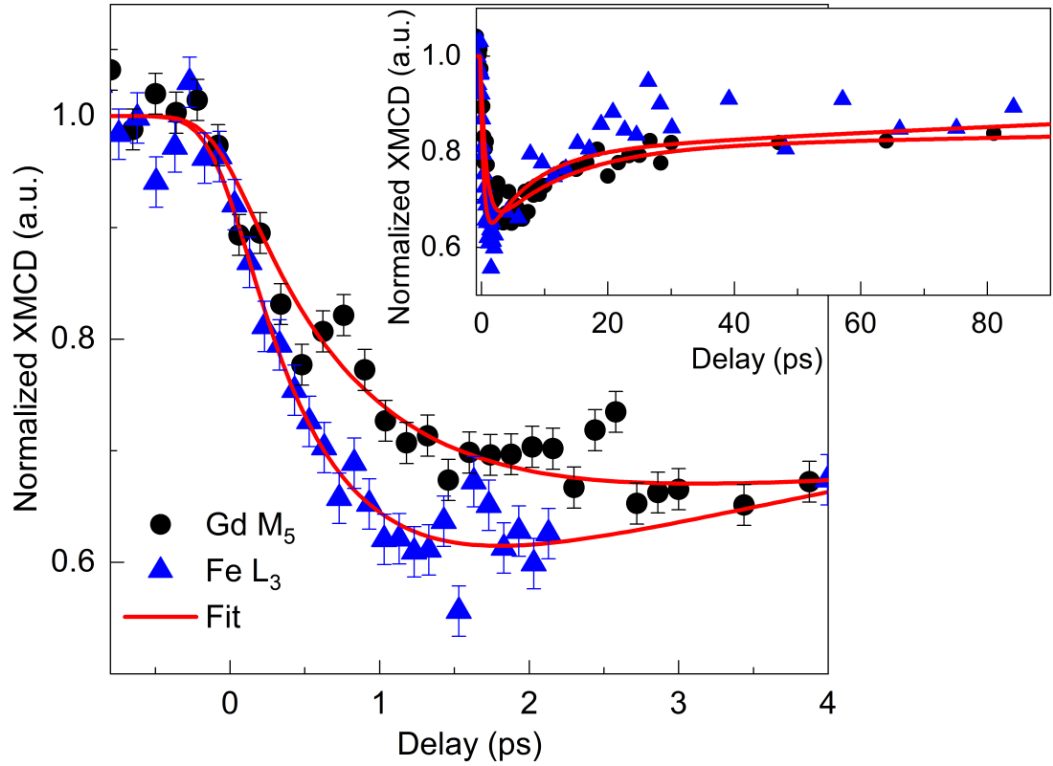

Figure 4. Dynamics induced by unpolarized hot electrons measured at Fe L3 and Gd M5 edges at incident IR laser fluence of 40mJ/cm$^2$ and simulated with absorbed fluence of 1.2 mJ/cm$^2$. The estimated error bars are the standard deviation of the experimental data with respect to the fitting function.

**E- IMPACT OF SIMULATION PARAMETERS ON MAGNETIZATION DYNAMICS**

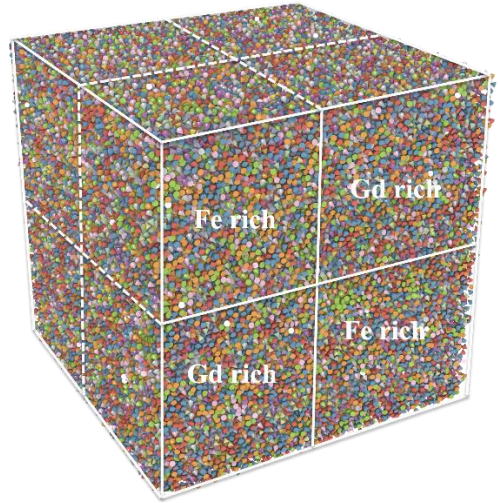

Figure 5. Schematic figure showing the heterogeneous amorphous samples of $Fe_{74}Gd_{26}$ with marked areas of areas richer in Gd or in Fe.

In our atomistic spin dynamics calculations, we use amorphous $Fe_{74}Gd_{23}$ with 1600 atoms in the simulation cell with periodic boundary conditions. Our simulation sample with areas rich in Gd or Fe is shown in Fig.5. The simulations parameters, such as hot electrons pulse, current density, Gilbert damping $α$, and electron-spin $G_{es}$, electron-phonon $G_{el}$, spin-lattice $G_{sl}$ coupling in three temperature models will impact the resulting magnetization dynamics, its amplitude, demagnetization, and remagnetization rates. Here, we demonstrate the influence of these parameters Figs.6-11. In particular, we start by varying hot electrons pulse used in our simulations on magnetization dynamics of iron and gadolinium in amorphous $Fe_{74}Gd_{26}$. In Fig.6, we present the impact of fluence for both negative and positive current directions. The shaded areas correspond to the values interval 2.8

– 3.6 mJ/cm$^2$ for iron and 1.0 – 1.4 mJ/cm$^2$ for gadolinium. Then, we add a similar analysis for electron-spin coupling $G_{es}$, as one can see in Fig.7, electron-phonon $G_{el}$ (Fig.8). In our simulations, presented in the main text, we do not consider spin-lattice coupling, as one can see in Fig.9 it is indeed not significant, especially for Gd, and leading mostly to slight change of the demagnetization amplitude. In addition, our simulations, presented in the main manuscript assume that the observed difference between parallel and antiparallel cases in the experiment is due to STT impact. Fig.10 shows our results are impacted by a change of STT values. One can observe, that, the difference between parallel and antiparallel cases is maintained, while STT is changed slightly. We demonstrate the impact of Gilbert damping in Fig.11. Finally, the impact of the pulse duration on magnetization dynamics is shown in Fig.12. In all simulations, we keep all parameters fixed and vary only one of them, e.g. laser fluence in Fig.6. The solid lines on all graphs corresponding to simulations presented by solid lines in Figs.5-6 of the main manuscript.

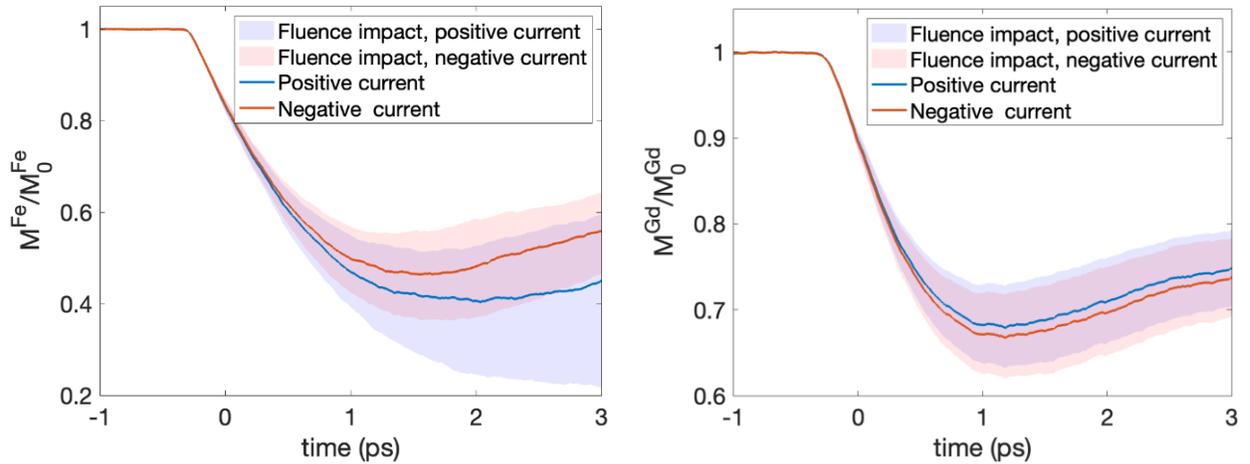

Figure 6. The impact of hot electrons pulse on magnetization dynamics of Fe (left) and Gd (right) in amorphous Fe$_{74}$Gd$_{26}$. Shaded areas demonstrate the fluence interval 2.8 –3.6 mJ/cm$^2$ for iron and 1.0 – 1.4 mJ/cm$^2$ for gadolinium. The blue/red shaded areas/lines correspond to the cases of either positive or negative current in STT.

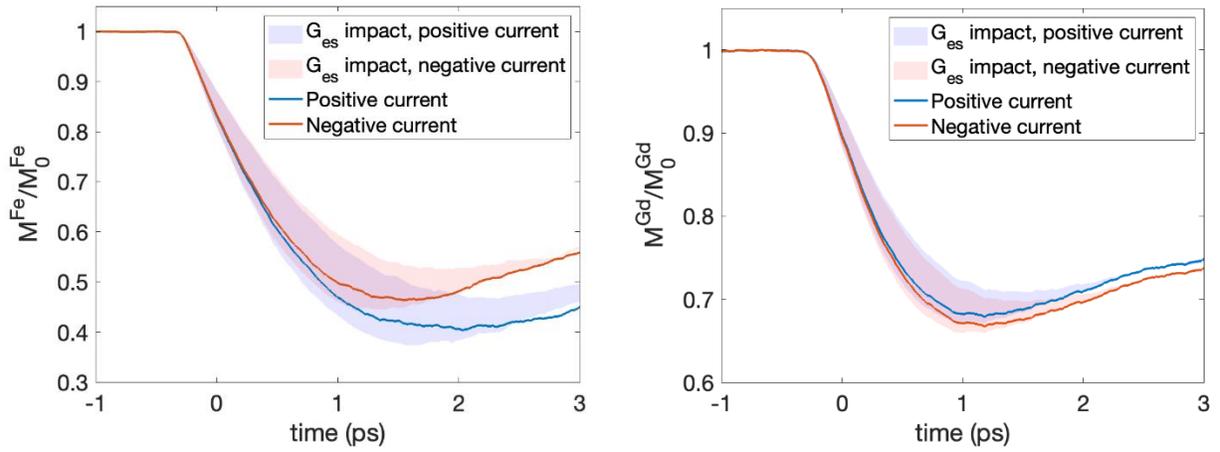

Figure 7. The impact of electron-spin coupling $G_{es}$ on magnetization dynamics of Fe (left) and Gd (right) in amorphous $Fe_{74}Gd_{26}$. Shaded areas demonstrate the $G_{es}$ interval $0.8 \times 10^{18} - 1.4 \times 10^{18}$ $W/m^3/K$. The blue/red shaded areas/lines correspond to the cases of either positive or negative current in STT.

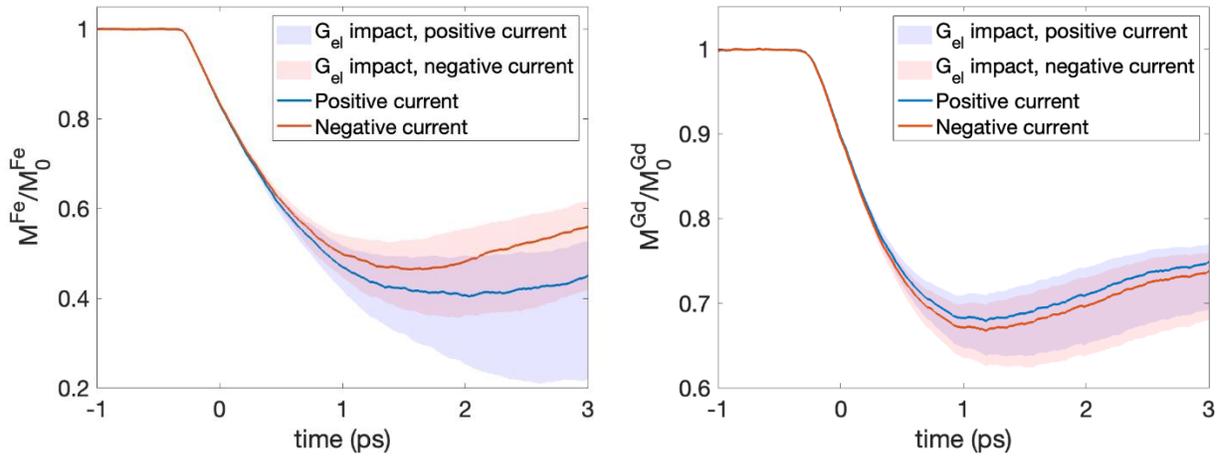

Figure 8. The impact of electron-phonon coupling $G_{el}$ on magnetization dynamics of Fe (left) and Gd (right) in amorphous $Fe_{74}Gd_{26}$. Shaded areas demonstrate the $G_{el}$ interval $4 \times 10^{17} - 1.2 \times 10^{18}$ $W/m^3/K$. The blue/red shaded areas/lines correspond to the cases of either positive or negative current in STT.

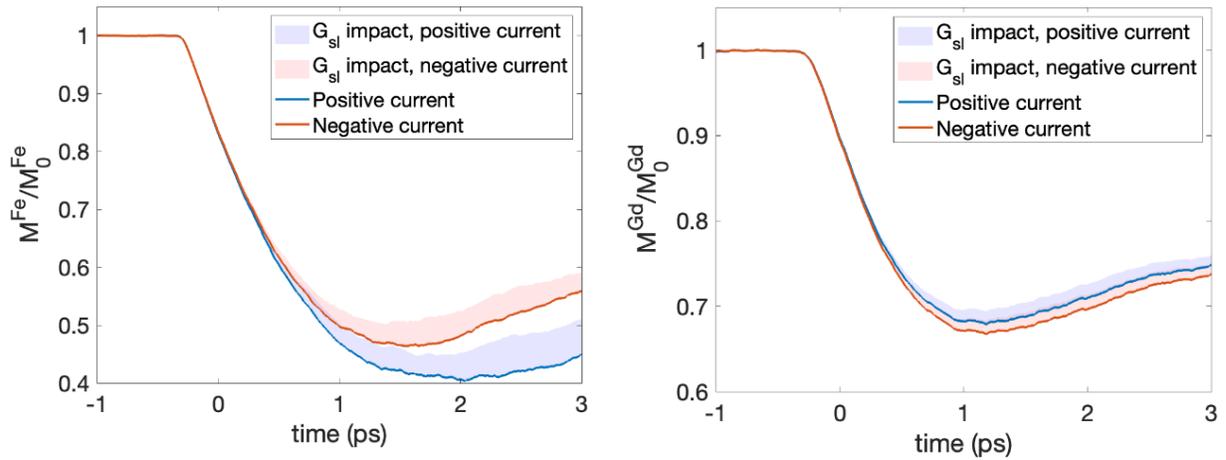

Figure 9. The impact of spin-lattice coupling $G_{sl}$ on magnetization dynamics of Fe (left) and Gd (right) in amorphous $Fe_{74}Gd_{26}$. Shaded areas demonstrate the $G_{el}$ interval $0.0 - 1 \times 10^{17}$ W/m$^3$/K. The blue/red shaded areas/lines correspond to the cases of either positive or negative current in STT.

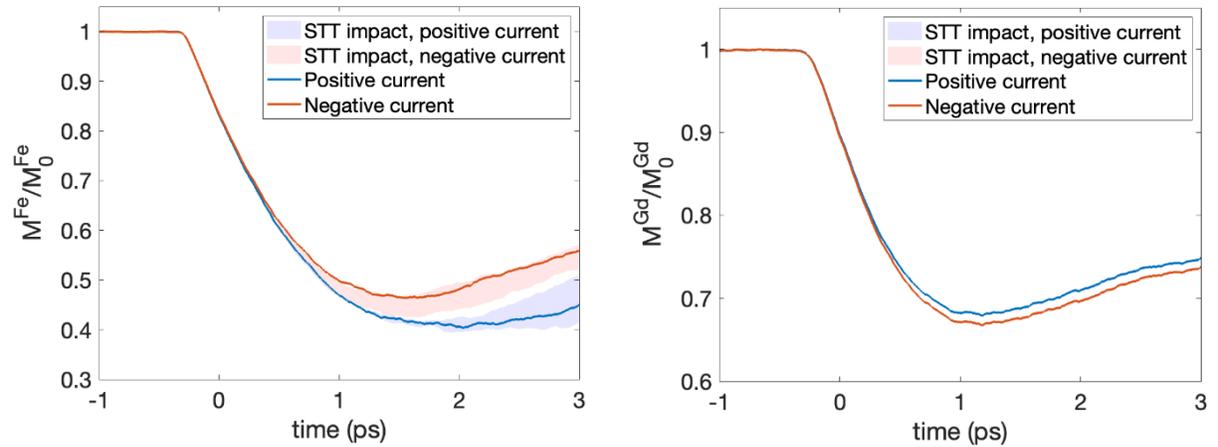

Figure 10. The impact of the strength of the STT on magnetization dynamics of Fe (left) and Gd (right) in amorphous $Fe_{74}Gd_{26}$. Shaded areas demonstrate the variation of the strength of the STT in the interval for +/- 25 % from the value given by the solid line. The current density interval for iron is $3.9 - 6.8 \times 10^{13}$ (assuming 50% polarization) or $1.9 - 3.4 \times 10^{13}$ (assuming 100% polarization). The current density interval for gadolinium is $1.4 - 2.5 \times 10^{13}$ (assuming 50% polarization) or $0.71 - 1.2 \times 10^{13}$ (assuming 100% polarization). The blue/red shaded areas/lines correspond to the cases

of either positive or negative current in STT.

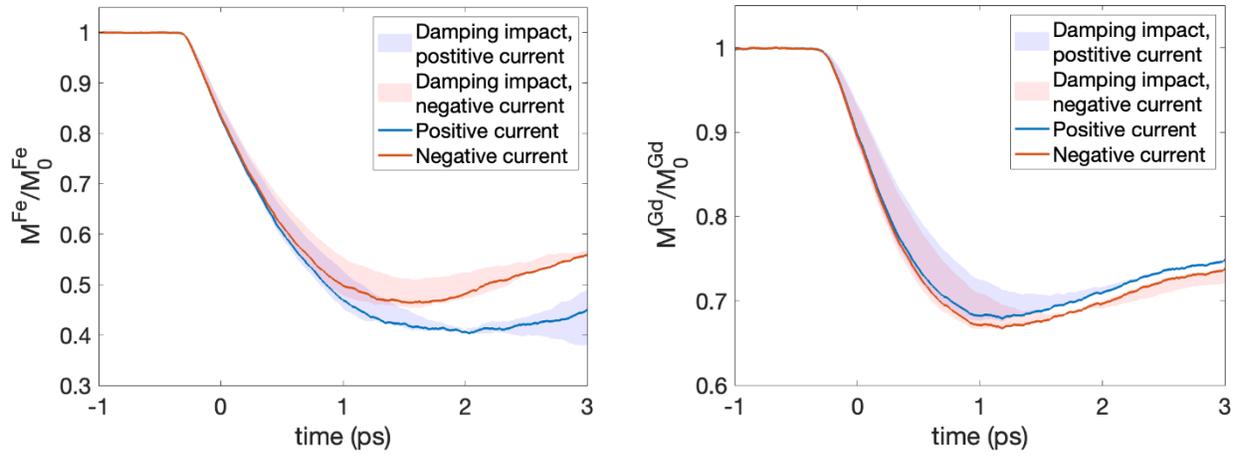

Figure 11. The impact of Gilbert damping *α* on magnetization dynamics of Fe (left) and Gd (right) in amorphous $Fe_{74}Gd_{26}$. Shaded areas demonstrate the *α* interval 0.05 – 0.12. The blue/red shaded areas/lines correspond to the cases of either positive or negative current in STT.

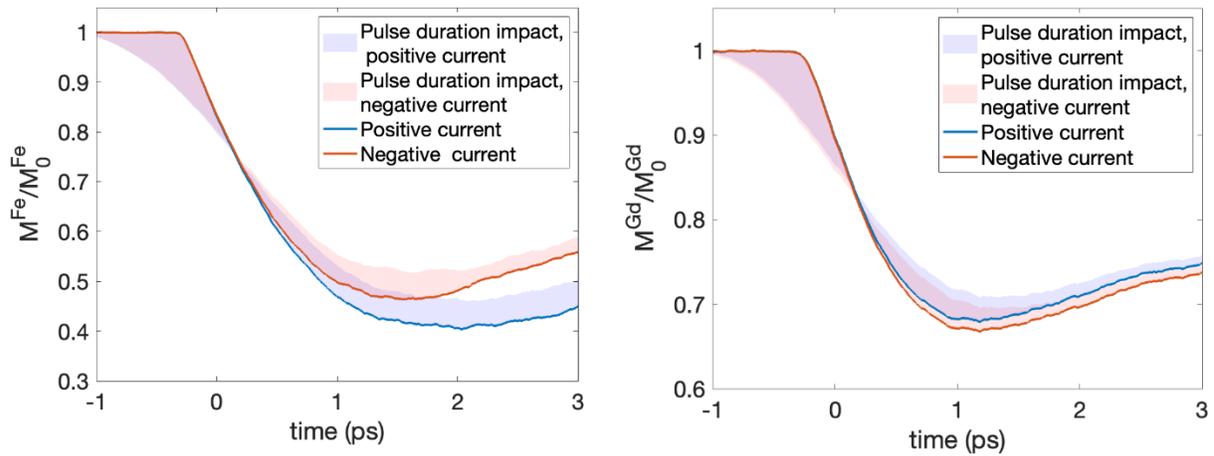

Figure 12. The impact of pulse duration on magnetization dynamics of Fe (left) and Gd (right) in amorphous $Fe_{74}Gd_{26}$. Shaded areas demonstrate a pulse duration interval 20 fs – 500 fs. The blue/red shaded areas/lines correspond to the cases of either positive or negative current in STT.

**PARAMETERS USED IN THE SIMULATIONS**

| Parameter | Value |
| --- | --- |
| Gilbert damping $\alpha$ | 0.1 |
| Electron-spin coupling $G_{es}$ | $1.2 \times 10^{18}$ W/m³/K. |
| Electron-phonon coupling $G_{el}$ | $8 \times 10^{17}$ W/m³/K. |
| Spin-lattice coupling $G_{sl}$ | 0.0 W/m³/K. |
| Temperature | 80 K |
| Current density (Fe, Fig. 5 and 7) | $2.6 \times 10^{13}$ A/m² (polarization 100%) |
| Current density (Gd, Fig.6) | $1 \times 10^{13}$ A/m² (polarization 100%) |
| Current density (Fe, Fig. 5 and 7) | $5.38 \times 10^{13}$ A/m² (polarization 50%) |
| Current density (Gd, Fig.6) | $2 \times 10^{13}$ A/m² (polarization 50%) |
| Current density (Fe, Fig. 5 and 7) | $2.6 \times 10^{14}$ A/m² (polarization 10%) |
| Current density (Gd, Fig.6) | $1 \times 10^{14}$ A/m² (polarization 10%) |

Table II. List of the parameters used in the simulations.

**F- Simulated element resolved demagnetization dynamics at high and low fluences (Fe3d versus Gd4f).**

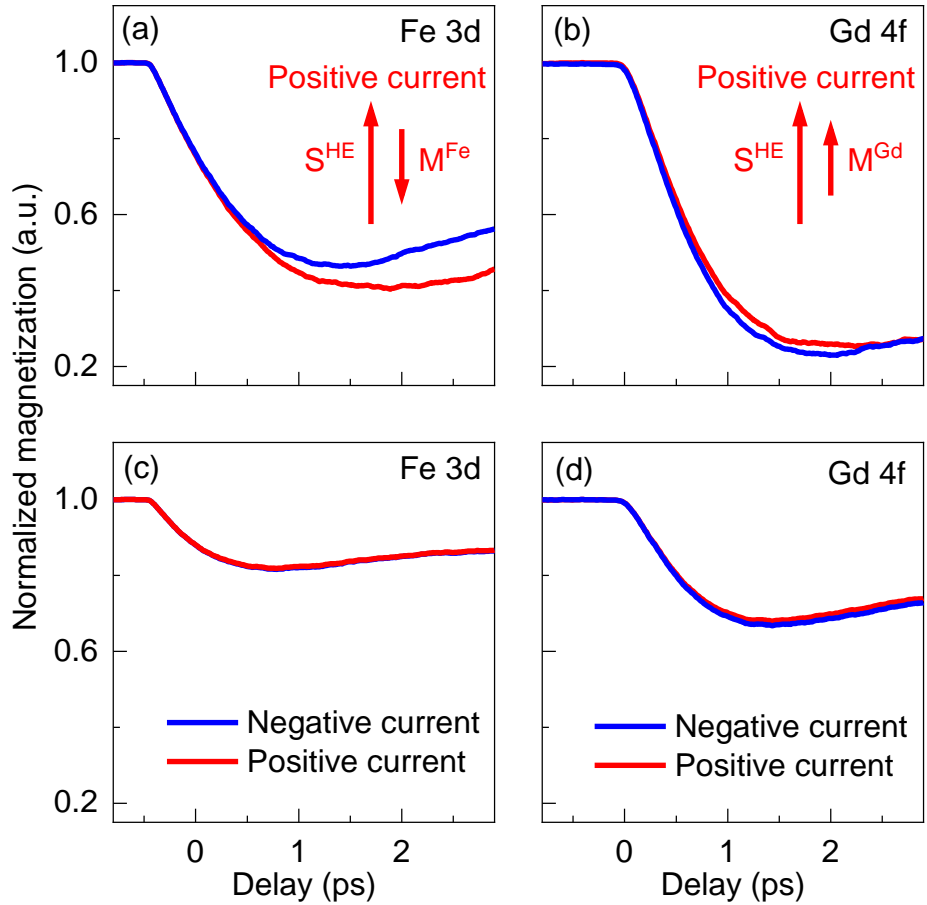

Figure 13. Simulated atomistic spin dynamics of Fe (a,c) and Gd (b,d) moments in $Fe_{74}Gd_{26}$ for the two absorbed fluence values (a,b) 3.2 mJ/cm$^2$ and (c,d) 1.2 mJ/cm$^2$. The AP configuration in the SPHE currents leads to deceleration of the 3dFe and acceleration of the 4fGd dynamics. The red and blue lines are simulations for opposite signs of spin polarized currents (positive and negative) combining heat-driven demagnetization dynamics with STT.

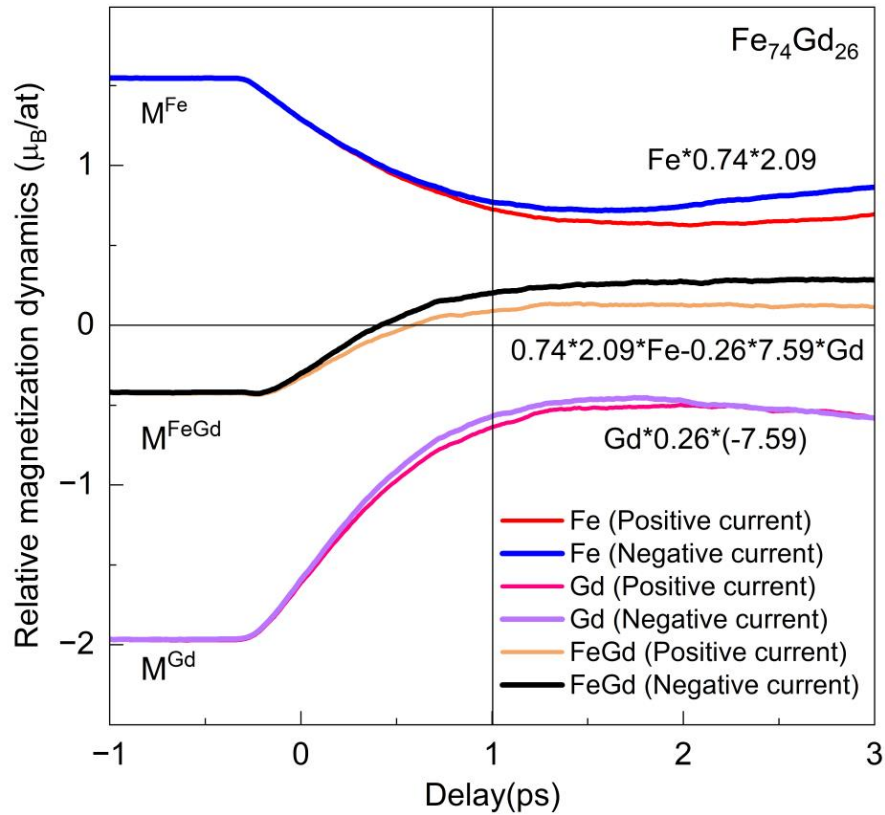

Figure 14. Simulated atomistic spin dynamics of Gd (continuous lines) moment in Fe$_{74}$Gd$_{26}$ for the absorbed fluence values of 3.2 mJ/cm$^2$ (120mJ/cm$^2$ incident laser fluence) using positive (red line - AP config) and negative (blue-violet lines - P config) spin polarized currents as given above in Figure 13. The orange and black curves correspond to the magnetization dynamics of the alloy Fe$_{74}$Gd$_{26}$ calculated by a linear combination of the Fe and Gd contributions and evidence an overall demagnetization changing his sign at ~0.4ps whereas the relative P/AP effect is large. For the ferrimagnetic alloy the net magnetic moment is strongly reduced so that the P/AP effect shows about 38 % at t= 1ps. The magnetic moment values are from Chimata et al. [5].